\documentclass[12pt,preprint]{aastex}

\def\etal{{et al.}}
\def\kms{\ {\rm km~s}$^{-1} $}

\begin{document}

\title{Jets or high velocity flows revealed in high-cadence spectrometer and imager co-observations?}

\author{M.S. Madjarska\altaffilmark{1}, J.G. Doyle\altaffilmark{2}, D. Innes\altaffilmark{1} 
and W. Curdt\altaffilmark{1}}

\altaffiltext{1}{Max-Planck-Institut f\"{u}r Sonnensystemforschung, Max-Planck-Str. 2,
37191 Katlenburg-Lindau, Germany}
\email{madjarska@mps.mpg.de}
\altaffiltext{2}{Armagh Observatory, College Hill, Armagh BT61 9DG, 
N. Ireland}

\shorttitle{Jets or high velocity flows}
\shortauthors{Madjarska et al.}

\begin{abstract}
 We report on active region  EUV dynamic events observed  simultaneously at high-cadence 
 with SUMER/SoHO and TRACE.  Although the features appear in the TRACE~Fe~{\sc ix/x}~171~\AA\ 
 images  as  jets seen in projection on the solar disk, the SUMER spectral line profiles 
 suggest that the plasma has been driven along a curved large scale magnetic structure, 
 a pre-existing loop.  The SUMER observations were carried out in spectral lines covering 
 a large temperature range from 10$^4$~K to 10$^6$~K. The spectral analysis revealed 
 that a sudden heating from an energy deposition is followed by a high velocity plasma flow.
  The Doppler velocities were found to be in the range from 90 to 160 \kms.
 The heating process has a duration which is below the SUMER exposure time of 25 {\rm s} 
 while the lifetime of the events is from 5 to 15 {\rm min}. The additional check on 
 soft X-ray Yohkoh  images shows that the features most probably reach 3 MK (X-ray) 
 temperatures. The spectroscopic analysis showed no existence of cold material during the events.

\end{abstract}


\section{Introduction} 
A large variety of jet-like phenomena are often observed in the solar atmosphere such 
as surges, spicules, sprays, Extreme-UltraViolet (EUV) and X-ray jets. X-ray jets 
(Shibata \etal\ 1992) were first identified in data obtained with the Soft X-ray 
Telescope (SXT) on Yohkoh (Tsuneta \etal\ 1991). They represent X-ray enhancements 
with an apparent collimated motion and were found to have a typical size of 5~$\times$~10$^3$ -- 
4~$\times$~10$^5$~{\rm km} and an apparent velocity of 30 to 300 \kms. Their kinetic 
energy is estimated  to be 10$^{25}$ -- 10$^{28}$~{\rm ergs}. Most of the jets were associated 
with small flares in large X-ray bright points or active regions. Shimojo \& Shibata (2000)
derived the physical parameters of X-ray jets and found temperatures from 3 to 8 MK (determined by 
using Yohkoh filter ratios) and densities of 0.7 -- 4.0 $\times$ 10$^9$~{\rm cm}$^{-3}$. 
It is strongly believed that they are produced by magnetic reconnection and represent the 
evaporation flow resulting from the reconnection heating. 

EUV jets were studied by Brekke (1999) in off-limb data from the Coronal Diagnostics
Spectrometer (CDS) and the Extreme-ultraviolet Imaging Telescope (EIT). From the 
CDS data it was found that the jet was emitting only at transition region temperatures 
showing Doppler shifts in the O~{\sc v} 629.73 \AA\ line up to -75 \kms. The event was 
also seen in the EIT Fe~{\sc xii}~195~\AA\ passband propagating with an apparent velocity 
of 180 \kms. The plasma seemed to be ejected along a large looped magnetic structure.
 Jets were also analysed in  on-disk data from 
the  Transition Region And Coronal Explorer (TRACE) taken in the 171~\AA\ and 
1216~\AA\ passbands by Alexander \& Fletcher (2000). In the 171~\AA\ channel the ejected 
plasma was seen both in emission and absorption which suggests that simultaneously highly 
collimated hot and cold  material was ejected along the magnetic field lines.  An EUV jet 
from a new emerging active region (a large Bright Point) was analysed in simultaneous 
TRACE, EIT and CDS data by Lin \etal\ (2006). The authors found the plasma jet to emit in a wide 
temperature range from 10\,000~K (He~{\sc i}) to 2.5~MK (Fe~{\sc xvi}, the upper 
temperature limit  of their observations).

H$_\alpha$ surges are often associated with EUV and X-ray emissions showing the co-existence of 
cool (H$_\alpha$) and hot ejections of plasma (Jiang \etal\ 2007 and references therein). 
Only recently, however, have the spatial and temporal relation of these emissions been 
studied in detail (Jian \etal\ 2007) during surge events in a plage area 
of an active region. The authors first observed the bright structures in TRACE 171 \AA\ 
followed by the cooler  H$_\alpha$ jet which they interpret as cooling of the hot 
plasma with a cooling time lasting about 6 -- 15 {\rm min}.

\section{Observations}

The events discussed here occurred in the active region NOAA 8558 on 1999 June 2. No 
flares were registered during the events. Simultaneous Solar Ultraviolet Measurements 
of Emitted Radiation (SUMER) telescope and TRACE observations were taken 
during several hours. The field-of-view (FOV) of the two instruments is shown in Figure~\ref{fig-1}. 
EIT Fe~{\sc xii}~195~\AA\ single  images for some of the events are also available, 
as well as a  few SXT images. In the  present letter only TRACE and SUMER data will be shown.

The SUMER spectrometer (Wilhelm \etal\ 1995, Lemaire \etal\ 1997) data were taken on 1999 June 2 
 starting at 09:17~UT and ending at 11:02~UT. A slit with a size of 0\farcs3 $\times$ 
120\arcsec\ was used exposing for 25~{\rm s} pointed at 
the plage area of the active region between two sunspots (Figure~\ref{fig-1}). Four spectral 
windows were telemetred, each with a size of 120 spatial $\times$ 50 spectral pixels.
 The spectral lines read out are shown in
Table~\ref{table:1}.  At the start of the observations the spectrometer was pointed at 
solar disk coordinates 
 xcen~=~-217~\arcsec (at 09:17~UT) and ycen~=~257\arcsec. Subsequently, the observations 
 were compensated for the solar rotation. The spectral analysis was made in 
 respect to a reference spectrum obtained by averaging over the entire dataset. 
 
The TRACE (Handy \etal\ 1999) data  were obtained in the Fe~{\sc ix/x}~171~\AA\ and 1600~\AA\ 
passbands starting at 09:00~UT and finishing at 11:30~UT on June 2, 1999.  
The integration time was 2.9~{\rm s} for the 171~\AA\ passband and 0.3~{\rm s} for 1600~\AA. 
 The 171~\AA\ channel cadence was 10~{\rm s} which increased to 15 {\rm s} when an image in the 
1600~\AA\ channel was taken. From 09:18:38~UT until 09:32:28~UT only observations in the 1600~\AA\ 
channel were taken.

The co-alignment of SUMER and TRACE observations (spatial resolution of 1.5\arcsec\ 
and 1\arcsec, respectively) was done by using SUMER raster observations,
taken just before the time series in the Si~{\sc ii}~1260.44~\AA\ line 
which  falls in the transmitted O~{\sc v} 629.73~\AA\ spectral window, and  TRACE~1600~\AA\ images.
 The SUMER raster was obtained with 5~{\rm s} exposure time and 0.37\arcsec\ increment.
The emission in the TRACE~1600~\AA\ passband  mainly comes from continuum emission, C~{\sc iv}, 
C~{\sc i},  and Fe~{\sc ii}. Note that the SUMER times mark the beginning of the exposures, 
while the TRACE times the end of the exposures.

\section{Features analysis: temporal evolution, velocity and temperature}

\begin{figure}
\epsscale{0.7}
\plotone{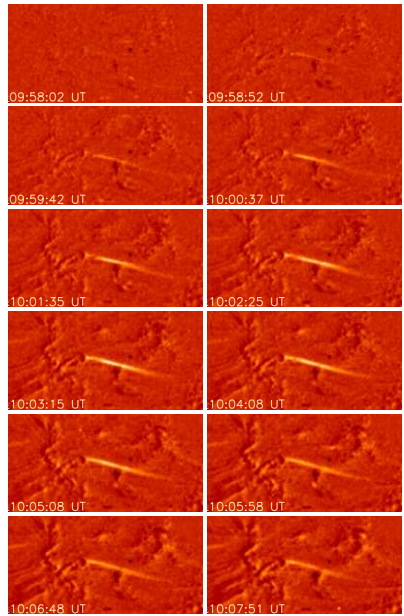}
\caption{TRACE 171 \AA\ image sequence of EV3. The dimenstions of the images are 100\arcsec 
$\times$ 50\arcsec\
 and were taken  approximately 50 {\rm s} apart.
\label{fig-2}}
\end{figure}

We identified three events in the SUMER data which we will call further in the text 
EV1, EV2 and EV3. Only EV3 was fully co-registered by SUMER and TRACE. Figure~{\ref{fig-2}} 
demonstrates through a sequence of difference images the appearance of EV3 as a jet-like event 
in the TRACE images. The difference images were obtained by substracting an image taken 
at 09:55:49~UT. The entire lifetime of EV1 and EV2 is only seen in the SUMER data, while 
TRACE registered only the fading phase of these features.

\begin{figure}
\epsscale{0.8}
\plottwo{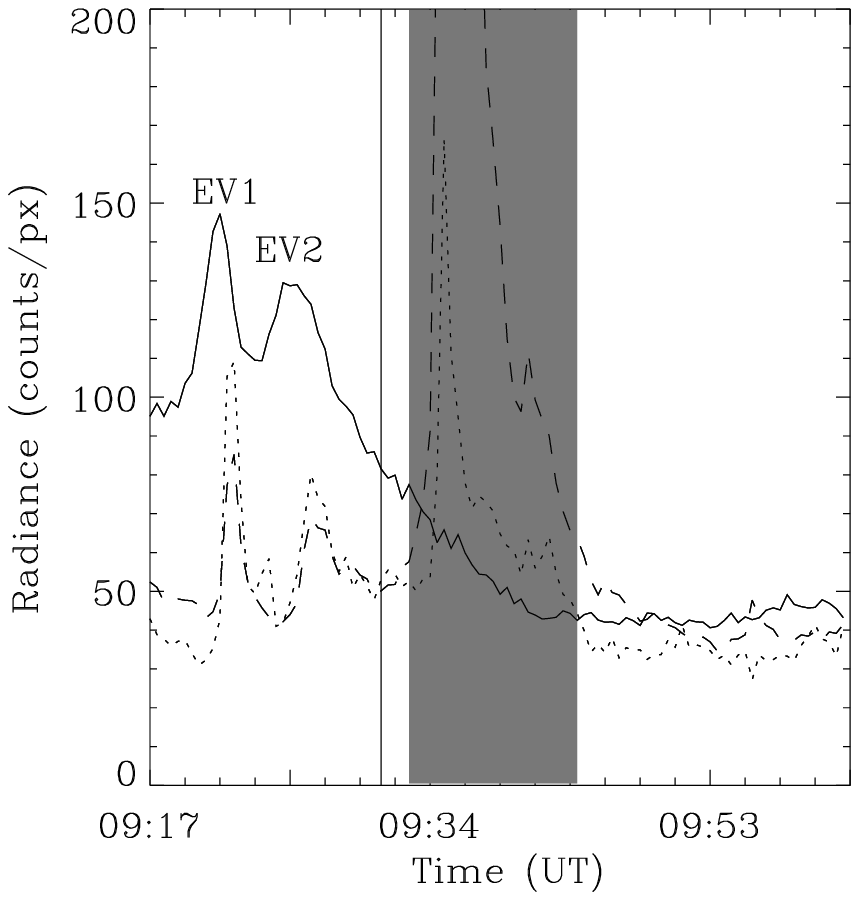}{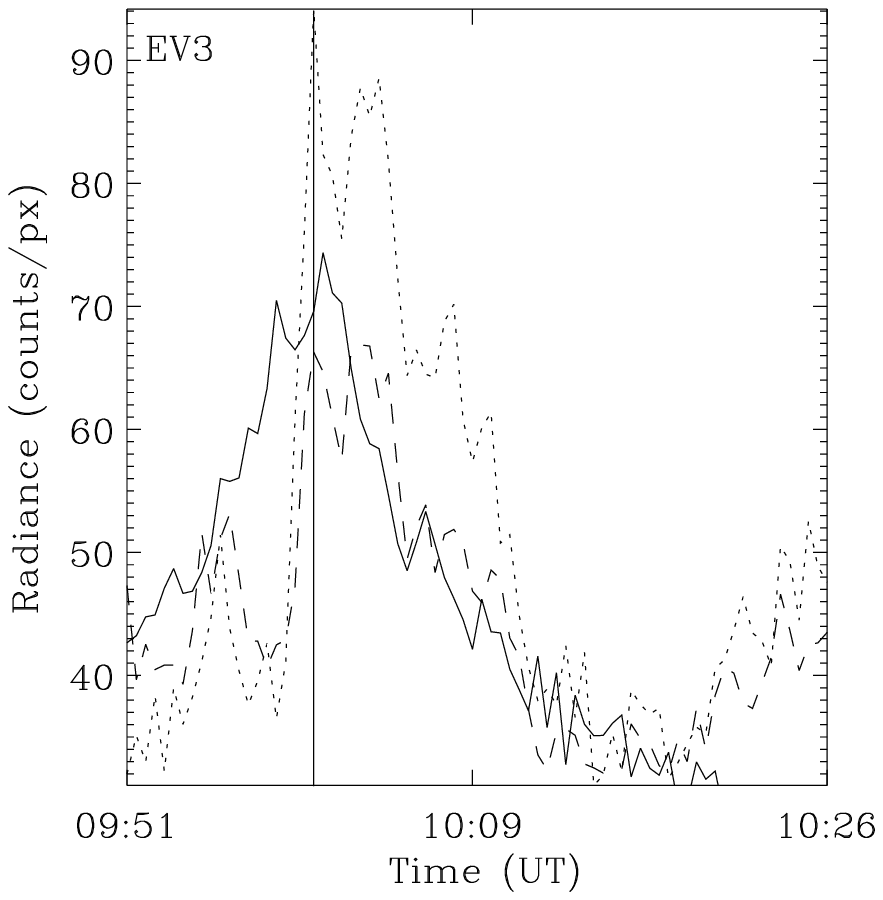}
\caption{{\bf Left:} The peak radiance (background substructed) averaged over 11 pixels along 
the SUMER slit obtained from a single Gauss fit during EV1 and EV2. The solid line corresponds to 
the Mg~{\sc x}~625~\AA\ line, dotted -- O~{\sc v}~629~\AA\ and dashed -- N~{\sc v}~1238~\AA. 
The vertical solid line shows the time of the restart of the TRACE~171~\AA\ image taking 
(see Section 2). The shaded area marks the time period during which another event appeared 
along the line-of-sight  unrelated to the features discussed here (See for details Madjarska 
\etal\ 2007). {\bf Right:} The peak radiance (background substructed) averaged over 7 pixels along the 
SUMER slit  during EV3. The linestyles have the same correspondence as in the left Figure.
\label{fig-3}}

\end{figure}

 EV1 started at 09:20:29~UT (12 min before the first  TRACE~171~\AA\ image after the observing gap,
 see Section 2) with a sudden increase in the emission of the Mg~{\sc x}~625~\AA\ line 
 (Figure~\ref{fig-3}, left) coupled with a Doppler shift of up to 30 \kms\ to the red. This 
 increase is seen over-imposed on the already relatively higher and red-shifted emission of 
 the Mg~{\sc x}~625~\AA\ line due to a pre-existing feature (not discussed in this paper). 
 Around 2 {\rm min} later, at 09:22:20~UT, the emission in O~{\sc v}~629~\AA\ and 
 N~{\sc v}~1238~\AA\ started to increase rapidly as well, reaching a maximum  around 75~{\rm s} 
 (three exposures) after the maximum in the Mg~{\sc x}~625~\AA\ line. A strong red-shift 
 is seen in the O~{\sc v}~629~\AA\ line (up to 110 \kms) and just a few \kms\ in 
 the N~{\sc v}~1238~\AA\ line. 

The  detailed study on the line profiles revealed that the feature started with a sudden large 
radiance increase of the rest component of the spectral lines and a red-shifted component. 
This suggests that an energy deposition took place followed by a collimated high velocity plasma 
flow. The heating process and the initial acceleration had a duration below the exposure time 
of 25~{\rm s}. It took around 75~{\rm s} for the three lines (Mg~{\sc x}, O~{\sc V} and N{\sc v}) 
to reach their maximum with the response in O~{\sc v}~629~\AA\ and N~{\sc v}~1238~\AA\, coming 
approximately 75 {\rm s} (three exposures) later. That was followed by a further acceleration of the 
flow and simultaneous decrease of the emission at the rest component in less 
than a minute. We clearly see the cooling of the event as a delay in the response in 
O~{\sc v}~629~\AA\ and N~{\sc v}~1238~\AA\ (Figures~\ref{fig-3} and 4). The feature had 
a stronger presence in the O~{\sc v}~629~\AA\ and Mg~{\sc x}~625~\AA\ lines and  a modest response 
in the N~{\sc v}~1238~\AA\ line with no signature at chromospheric 
temperatures, indicating a high electron density (Doyle \etal\ 2006a). Its lifetime was 
around 10 {\rm min}. The Doppler shifts in the O~{\sc v}~629~\AA\
 line are in the range from 90 to 160 \kms\ derived from a double Gauss fit.

\begin{figure}
\plottwo{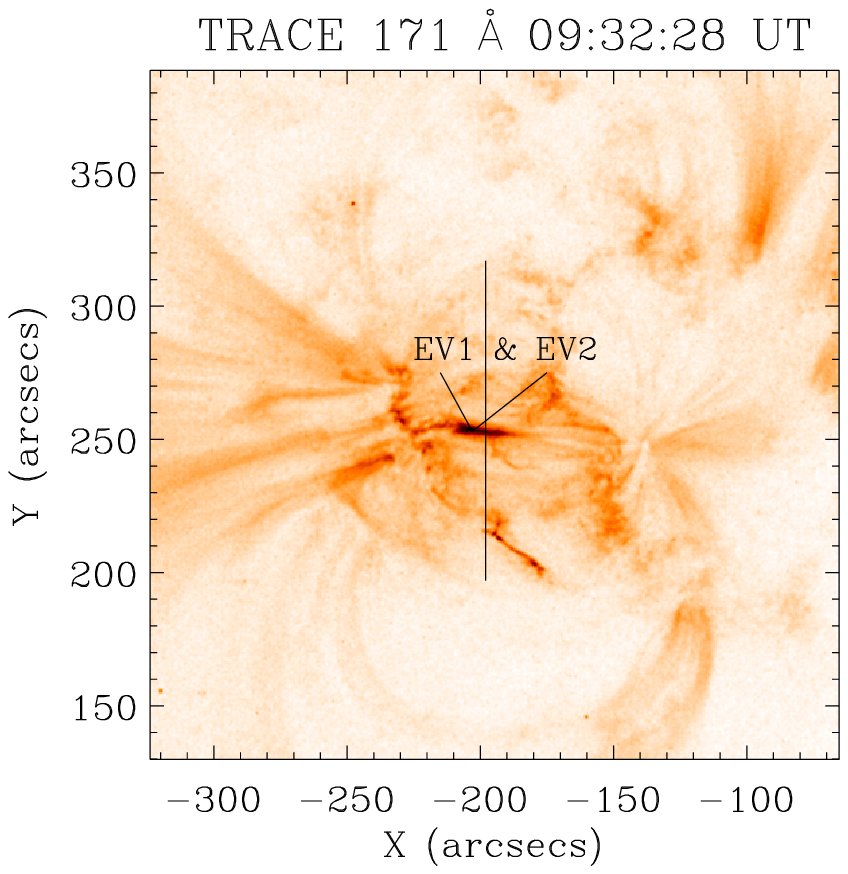}{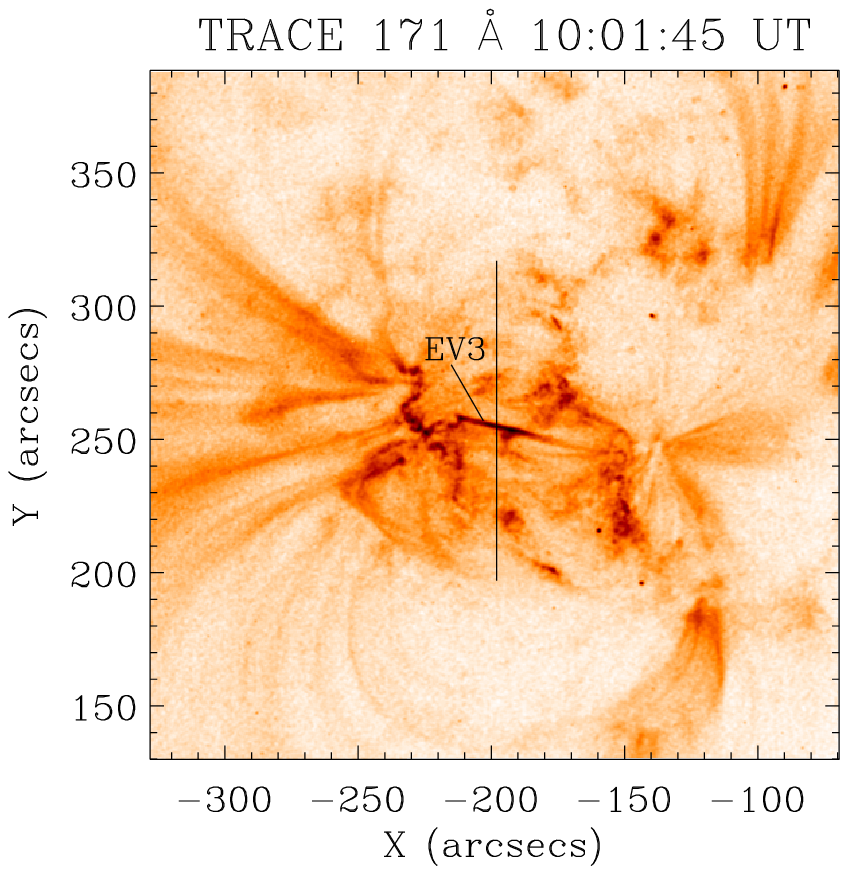}
\caption{Colour table reversed TRACE  Fe~{\sc ix}~171~\AA\ images obtained on 1999 June 2.
  The images are derotated to a reference time. The overplotted vertical line shows the 
  SUMER slit position.}
\label{fig-1}
\end{figure}

\begin{figure}
\epsscale{0.5}
\plotone{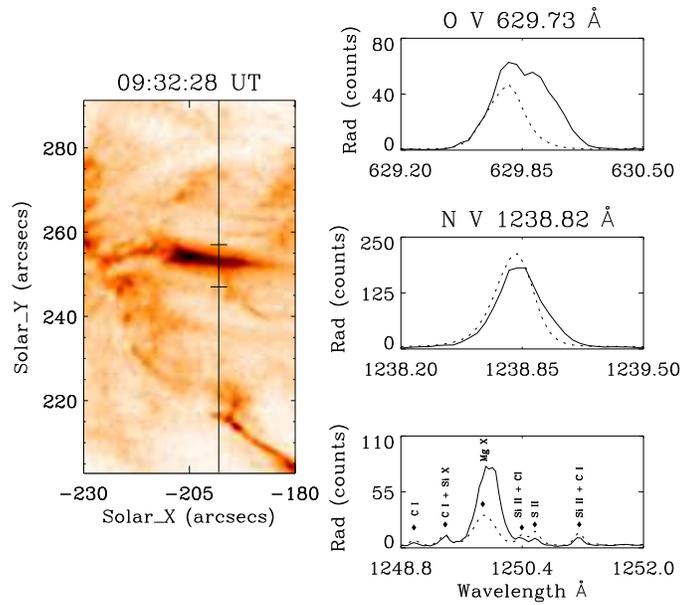}
\caption{{\bf Left:} TRACE 171 \AA\ color table reversed image showing the first two events 
-- EV1 \& EV2. The time of the image corresponds to the vertical solid line on  Figure~\ref{fig-2}, 
left. The vertical line marks the position of the SUMER slit, while the two horizontal lines 
outline the slit position which was analysed in the SUMER data.	{\bf Right:} The SUMER spectral 
line profiles at 09:32:06~UT. The dotted lines show the reference spectra. 
\label{fig-4}}
\end{figure}

EV 2 started 5 {\rm min} after EV1 around 09:25:05~UT appearing along the SUMER slit just 
above EV2. The maximum emission in the Mg~{\sc x}~625~\AA\ line lasted during 3 exposures 
and the delay  in the response of the transition region lines was again around 2 {\rm min}. 
The first TRACE~171~\AA\ image at 09:32:28~UT revealed a jet-like event (see Figure~\ref{fig-1}, 
left). The time of this image corresponds to the decaying phase of the features seen in the 
SUMER data (Figure~\ref{fig-3}, left). In the SUMER data at the start of EV2, it was still 
possible to separate both features along the slit. During the decaying phase, however, the 
events  appeared as a single feature along the SUMER slit, and that is how they were seen 
in the first TRACE image at 09:32:28~UT  (see Figure~\ref{fig-1}, left). 
In Figure~\ref{fig-4} the TRACE~171~\AA\  image  taken at 09:32:28~UT and the time 
corresponding SUMER spectral line profiles during the events (EV1 \& EV2) are shown. The 
features appeared as a radiance increase of the rest component of the O~{\sc v}~629~\AA\ 
line (Figure~\ref{fig-3}) and a strong red-shifted component (Figure~\ref{fig-4}). 
The same is observed in the Mg~{\sc x}~625~\AA\ line although the Doppler shift is smaller. 
No change in the emission of the chromospheric lines (Table 1) was observed. The width of 
EV2 is around 3\arcsec\ derived from its projection along the SUMER slit.

\begin{figure}
\epsscale{0.5}
\plotone{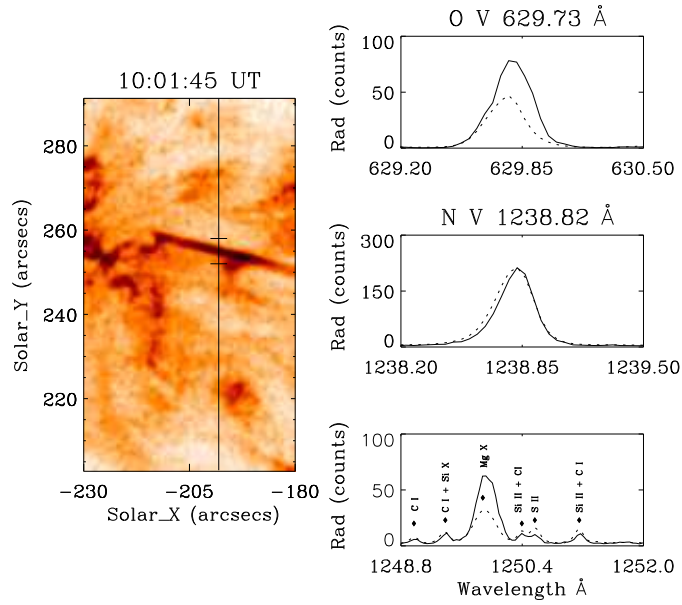}
\caption{{\bf Left:} TRACE 171 \AA\ color table reversed image showing EV3. 
The time of the image corresponds to the vertical solid line on  Figure~\ref{fig-2}, 
right. The vertical line marks the position of the SUMER slit, while the two horizontal 
lines outline the slit position which was analysed in the SUMER data. {\bf Right:} 
The spectral line profiles at 10:01:16~UT. The dotted lines show the reference spectra.
\label{fig-5}}
\end{figure}

EV3 was registered during the observations starting at 09:59:52~UT with a width of 
$\approx$3\arcsec\ (Figure~\ref{fig-1}, right). The temporal evolution of the amplitude of the 
radiance derived from a single Gauss fit in all three spectral lines can be seen 
in Figure~\ref{fig-3} (right). In Figure~\ref{fig-5} a TRACE 171 \AA\ image and the time 
corresponding SUMER line profiles  during EV3 are shown. The event showed the same 
temporal behaviour as the events described above and had a duration of around 15~{\rm min}. 
As for EV1 and EV2  no response was found in the chromospheric emission.

In order to find out with more precision whether cold plasma exist during the events 
we studied the ratio of the continuum emission of the SUMER first and second 
order radiation. The  first order emission ($\sim$1238 \AA) is above the H Lyman limit, 
while the second order ($\sim$629~\AA) is below. Second order emission will be absorbed 
by H~{\sc i} along the line-of-sight and the first order emission not (for details on the 
method see Innes \etal\ 2003). Therefore, the second- to first order ratio would show a 
decrease if cold material existed along the line-of-sight. The ratio did not show any 
changes which  indicated that emission at low temperatures was not present.  During all 
three events brightenings were observed in SXT and EIT Fe~{\sc xii}~195 \AA\ data. 
Such brightenings were often observed in SXT data and their energetics is studied 
in detail by  Shimizu (1995).

\section{Discussion}
  This letter presents, to our knowledge for the first time, EUV transient features in 
an active region identified and analysed in  on-disk SUMER  data and simultaneously 
obtained TRACE images. These instruments provide data at highest existing 
(1.5\arcsec and 1\arcsec) spatial and 2 \kms\ spectral resolution (SUMER). Additionally, 
the combination of spectrometer and imager data obtained at high cadence (25~{\rm s} and below) 
permitted the temporal and spatial evolution, velocities and especially temperatures of EUV 
active region transients to be derived with the highest possible precision. Three dynamic 
events were studied in spectral  lines covering a temperature range from 10$^4$ to 10$^6$~K. 
All three features  showed strong red-shifted emission  in the  O~{\sc v}~629~\AA\  and 
Mg~{\sc x}~625~\AA\ lines  suggesting high velocity flows which propagate in a direction 
away from the observer, i.e. towards the solar surface. Considering the 
magnetic fields structure  of the active region field by the loops seen in TRACE~171 \AA, we 
suggest that the features although appearing with a jet-like structure in TRACE~171 \AA\ images 
may rather represent a high velocity flow driven along a curved magnetic field, most 
probably a pre-existing loop. No signature of the events was found at chromospheric 
temperatures. Both EIT/Fe~{\sc xii}~195~\AA\ and Yohkoh/SXT showed brightenings in 
a pixel row indicating the presence of a 1 to 3 MK plasma during the transients.  The lower 
resolution of these instuments ($\approx$6\arcsec) in comparison to TRACE (1\arcsec) do not permit 
the events to be identified as jets.
The response in the transition region lines is delayed by  around 2~{\rm min} in respect 
to the coronal line suggesting cooling of the events. In the future hydrodynamic numerical 
simulations (see for detail Doyle \etal\ 2006b) will be performed  with the results 
converted into observable quantities to be then compared with the present data. 
We believe that the capabilities of  the Hinode mission will bring  a better 
understanding on these features and, more important, the physical mechanism behind them.
 
\acknowledgments
The SUMER project is financially supported by DLR, CNES, NASA, and PRODEX. 
Armagh Observatory's research is grant-aided by the N. Ireland Dept. of
Culture, Arts \& Leisure.

Facilities:\facility{SUMER/SoHO},
\facility{TRACE},
\facility{Yohkoh/SXT},
\facility{EIT/SoHO}

\clearpage

\begin{table}
\centering
\caption{The observed spectral lines.\label{table:1}}

\begin{tabular}{c c c c c}
\hline\hline
Ion & $\lambda$(\AA) & log(T)$_{max}$ & Comment\\
\hline

N~{\sc v} & 1238.82 & 5.3 & \\
C~{\sc i} & 1248.00 & 4.0 & \\
& 1248.88 & & blend\\
C~{\sc i} & 1249.00 & 4.0 &  \\
O~{\sc iv}/2 & 1249.24 & 5.2&  blend\\ 
Si~{\sc x}/2 & 1249.40 & 6.1 & blend\\
C~{\sc i} & 1249.41 & 4.0 & \\     
Mg~{\sc x}/2 & 1249.90 & 6.1 & \\
O~{\sc iv}/2 &1250.25&5.2& blend\\
Si~{\sc ii} & 1250.09&4.1 &  \\
Si~{\sc ii} & 1250.41 &4.1 & \\
C~{\sc i} & 1250.42 & 4.0 & blend\\
S~{\sc ii} & 1250.58 &4.2 &\\
Si~{\sc ii} & 1251.16 &4.1 & \\
C~{\sc i} & 1251.17 &4.0 & blend\\
O~{\sc iv}/2&1251.70&5.2&\\
Si~{\sc i} & 1258.78 &4.1 &\\
S~{\sc ii} & 1259.53 &4.2 &blend\\
O~{\sc v}/2&1259.54&5.4&&  \\
Si~{\sc ii}&1260.44&4.1&\\
\hline
\tablenotetext{}{Note: The expression /2 means that the spectral line was observed 
in second order. The comment `blend' means that the spectral line is 
 blending a close by line. The line formation temperatures are taken from CHIANTI v5.0 
 using the Mazzota \etal\ (1998) ionization equilibrium. }
\end{tabular}
\end{table}

\clearpage

\end{document}